# : Comparative Study of Lycopene Encapsulation Efficiency in Polycapprolactone Vs Poly Lactic Co-glycolic Acid

*Mohammad Anwar Ul Alam and Lamin S. Kassama,*
*Department of Food and Animal Science, Alabama A & M University, Normal, Huntsville, AL-35762, USA*

**ABSTRACT.** Lycopene contributes to the red-colored pigmentation of fruits and vegetables, and it is a fat-soluble carotenoid with antioxidant properties. Epidemiological studies have shown the significant health benefits associated to the consumption of lycopene rich foods, because of the anti-cancer properties. Degradative losses of lycopene during processing is a grave concern, hence encapsulation provides a remedy. The objective of this study is to evaluate the encapsulation efficiency of two biodegradable polymers (PLGA and PCL) as for controlled release of lycopene in the gastrointestinal (GI) system. The nanoparticles (NP) were synthesized by emulsion evaporation method and physicochemical properties was determined using a Dynamic Light Scattering spectroscopy. The results show the hydrodynamic diameter of the lycopene NP synthesized in PCL (200 mg) and 3500 mg surfactant and sonicated for 15 min was 79.23±0.85 nm (Lowest). PLGA (500 mg) and 500 mg surfactant with 15 min sonication was observed to have the lowest NP diameter (108.2±2.66 nm) among the others. Significant difference result found in PDI value (0.12±0.07) when PCL of 200mg dissolve in 3500 mg of surfactant. On the other hands the zeta potential values were much smaller in case of PCL NP ranged between -1.3±0.046 and -4.21±0.08 mV compared to the PLGA NP -72.36±2.17 to 107.66±3.15 mV in all experiments. Thus, NP synthesized with PCL and surfactant provide a smaller sized nano-solution than PLGA and surfactant. As the degradation rate for PCL is lower than PLGA so PCL can be considered as a potential biodegradable polymer than PLGA to encapsulate lycopene.

## Introduction:

Superoxide anion, hydrogen peroxide, hydroxyl, acyl and alkyloxy radical are highly reactive oxygen species can induce cell disorder by acting on their protein, lipids and DNA. Epidemiological studies by Gutteridge & Halliwell (2000); Halliwell (1996); Volko et al. (2007) have shown evidence that reactive oxygen species contributes to the development of artherosclarosis, repurfusion injury, cataractogenesis, rheumatoid arthritis, inflammatory disorders, cancer and ageing process. According to Ferreira et al. (2009) and Lopez et al. (2007) that plant based antioxidant (Lycopene, beta carotene, alpha tocopherol, ascorbic acid and polyphenols and flavonoids) can inhibit degenerative diseases with the free radical scavenging activity.

Lycopene is a fat-soluble carotenoid that act as an antioxidant after absorption through the intestine. Suwanaruang (2016) reported that lycopene contributes to the red pigmentation present in tomatoes, watermelon, pink guava and red bell pepper. The presence of lycopene in tomatoes increased it potential with highest free radical scavenging activity to reduce the capacity of singlet oxygen and its ability to trap peroxyl radical. Its antioxidant capacity is twice higher that beta carotene and ten times higher than alpha tocopherol. However, its activity is compromised by oxidation, isomerization, and degradation in processed foods during storage for an extended period. In their opinion, Calvo and Santagapita, (2017) indicated that the encapsulation of bioactive compound is a remarkable technique to prevent and preserve bioactive compound from degradation during processing and hence long-term storability. Reis et al. (2009); Sun and Deng (2005); Akagi et al. (2010); Jiang et al. (2009) have shown the need and scientific interest in developing a polymer based nano-encapsulation technique in the biomedical field for potential application in controlled release and targeted delivery of drugs, and bioaccessibility bioactive compounds in foods.

Guilherme et al. (2014) identified Polycaprolactone (PCL) which is a semi-crystalline aliphatic polyester for tissue engineering and bone tissue repair due to its biodegradability and non-toxigenic characteristics in the biomedical field. It behaves like plastic at normal temperature and pressure with a glass transition temperature ($T_g$) between -60 and $-10°C$ based on its molecular weight. Based on its biological and physicochemical properties, PCL has immense potential in the encapsulation of bioactive compound for controlled release.

PLGA is considered as a successful carrier as its metabolites are endogenous and are metabolized in the body through Krebs cycle by producing minimum systemic toxicity during the drug delivery and biomaterial application. Poly (lactic-co-glycolic acid) (PLGA) is a polymer, when hydrolyzed produce two metabolites such as lactic acid and glycolic acid. It is also approved by FDA and European Medicine Agency (EMA) for various drug delivery applications in human systems. It is available commercially with different molecular weights and copolymer compositions. Based on the molecular weight and copolymer ratio its degradation time can vary from several months to several years. Danhier et al. (2012) have shown that depending on its monomer ratio PLGA is identical in drug delivery application e.g. PLGA 50:50 consisting of 50% lactic acid and 50% glycolic acid. Hence, the main objective of this study is to identify the encapsulation efficiency of lycopene in PCL and PLGA by solvent emulsion evaporation technique.

## Materials and Methods:

**Chemicals:**

Lycopene ($M_w$ = 536.89) with HPLC grade supplied from Indofine Chemical Company Inc (Lot No: 1707841). Poly (D, L-lactic-co-glycolic) acid (PLGA) with a copolymer ratio of 1:1 [lactide/glycolide] and a $M_w$ of 10,000 was from Polyscitech (Lot: 60225ELH-A). Dimethylamino borane ($M_w$ =58.92) use as surfactant in encapsulating lycopene by PLGA was from Oakwood Chemical (Lot No: D14H). Poly-epsilon-caprolactone (PCL), polyvinyl alcohol and dichloromethene obtained from Scientific Polymer Products Inc, HACH Company and ACROS Organic respectively. All other chemicals were from various commercially approved chemical suppliers with a purity of analytical grade or better.

**Preparation of Lycopene loaded PCL nanoparticles:**

Nanoparticle synthesis was done by two steps process. In the first step, different concentrations ranging from 100 mg to 500 mg of PCL was dissolved in organic solvent, dichloromethane at room temperature under magnetic stirring. After homogenization, lycopene was introduced into the homogenized solution. In the second step different concentration of polyvinyl alcohol was introduced in 60 ml of distilled-deionized water at room temperature. Water was added into the organic solution and stirred for two minutes. After that the emulsion was formed by subjecting the solution to Ultrasonication (Sonicator 3000 Ultrasonic liquid processor, Church Hill, CT., U.S.A.) for certain specified time period. Organic solvent was evaporated using a rotary vacuum evaporator (RE 301, Yamato Scientific Co. Ltd., Tokyo, Japan) to recover the concentrated nanoparticle solution.

**Preparation of Lycopene loaded PLGA nanoparticles:**

Nanoparticle synthesis using by the PLGA loaded in lycopene was done in two steps emulsion evaporation technique. In the first step different concentrations of PLGA was titrated in ethyl acetate at room temperature under magnetic stirring. After uniform mixing, lycopene was added into that homogenized solution and mixed by stirring again. In the following step, dimethylamin boren was dissolved in specified amount of deionized water at room temperature. After that the emulsion was formed by running the sample through Sonicator (Sonicator 3000 Ultrasonic liquid processor, Church Hill, CT., U.S.A.). Ethyl acetate was evaporated from the nanoparticle solution with a rotary evaporator (RE 301, Yamato Scientific Co. Ltd., Tokyo, Japan).

**Determination of Physicochemical Characteristics**:

Hydrodynamic diameter, polydispersity index, zeta potential and conductivity will be measured by using a Dynamic Light Scattering spectroscopy Malvern Zetasizer Nano series (Nano ZS90, Malvern Instruments Ltd., Worcestershire, UK). The hydrodynamic diameter expressed as nano-meter (nm) and zeta potential measured as millivolt (mV) where PDI was a unit-less, hence values were determined and used in the characterization of their respective physical properties.

# Results and Discussion:

The biodegradable polymer PCL was used to encapsulate lycopene and surfactant (PVA) was used. Homogenization technique was by sonication. Hence, no significant ($p < 0.05$) change was found in between the concentration of PCL and hydrodynamic diameter (nm) of nanoparticle. When lycopene nanoparticle was synthesized in 200 mg of PCL and 3500 mg of PVA (surfactant agent) for at 15 min sonication time, hydrodynamic diameter found to be the lowest (79.23±0.85 nm) among the other experimental conditions as shown in Figure 1. Subsequently, PAV (3000 mg) both sonicated for 15 and 20 min, hydrodynamic diameter was found to fluctuate between 150 nm to slightly below 300 nm at different PCL concentrations. However, hydrodynamic diameter were found to be slightly higher when 100 mg of PCL and 2500 mg of PVA was used at 20 min sonication time but found to be constant (slightly above 200 nm) at 300 , 400 and 500 mg. One study of Guilherme (2014) evolved that nanoparticle size was 82 nm when PCL was used to encapsulate progesterone using by supercritical $CO_2$. However, when conventional method was used the hydrodynamic diameter of NP found to be 174 nm.

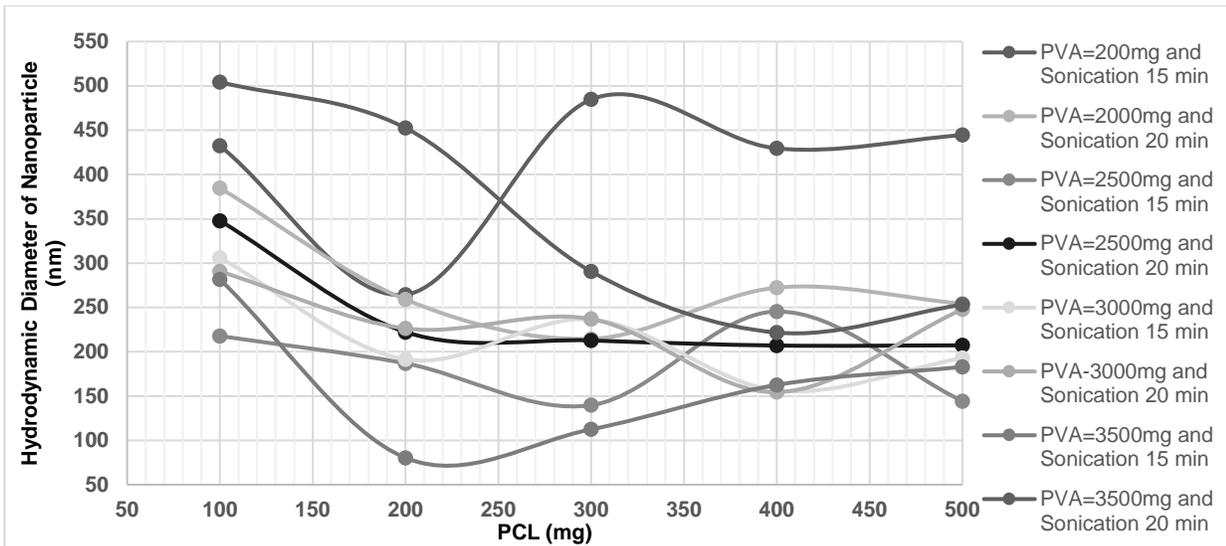

**Figure-1: The hydrodynamic diameter profile of NP in different PCL concentrations**

Figure 2 shows the hydrodynamic diameter profile of lycopene NP in different PLGA concentration. The lowest hydrodynamic diameter found to be 108.2±2.66 nm when 400 mg of PLGA and 500 mg of DMAB used (at 15 min sonication time).

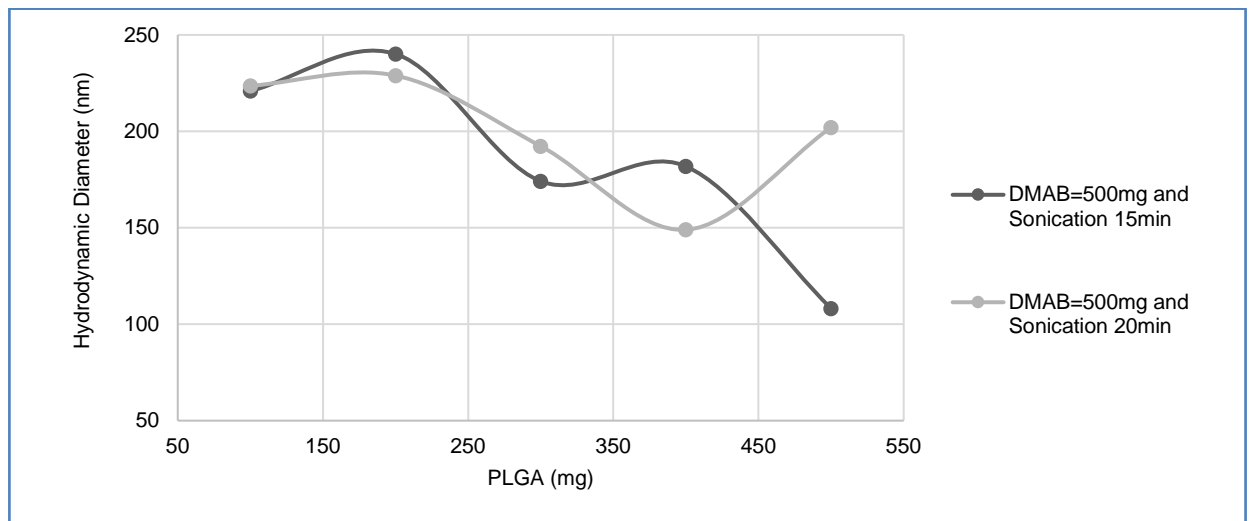

**Figure-2: The hydrodynamic diameter profiles of NP in different PLGA concentrations**

Bahari and Hamishehkar (2016); Bertrand and Lerpux (2012) stated that blood vessels do not permit the transportation of NP through the capillaries to the kidney and the lung and heart if the diameter is within the range of 100-150 nm. However, to obtain a successful delivery system targeting the bone marrow, spleen and the liver sinusoids, the NP size should range between 20-100 nm (Townsley et al., 1988). In another study Gratton et al., (2008) mentioned that NP in the size range of 100-150 nm have about 8-9 times greater capacity to target tumor cells compared to size range of 3-5 μm.

The relationship between nanoparticle size and percent intensity distribution as shown in Figure 3 appears to be normal. Hence, the size distribution (Figure 3) significantly ($p < 0.05$) is evident that particle sizes are homogeneous.

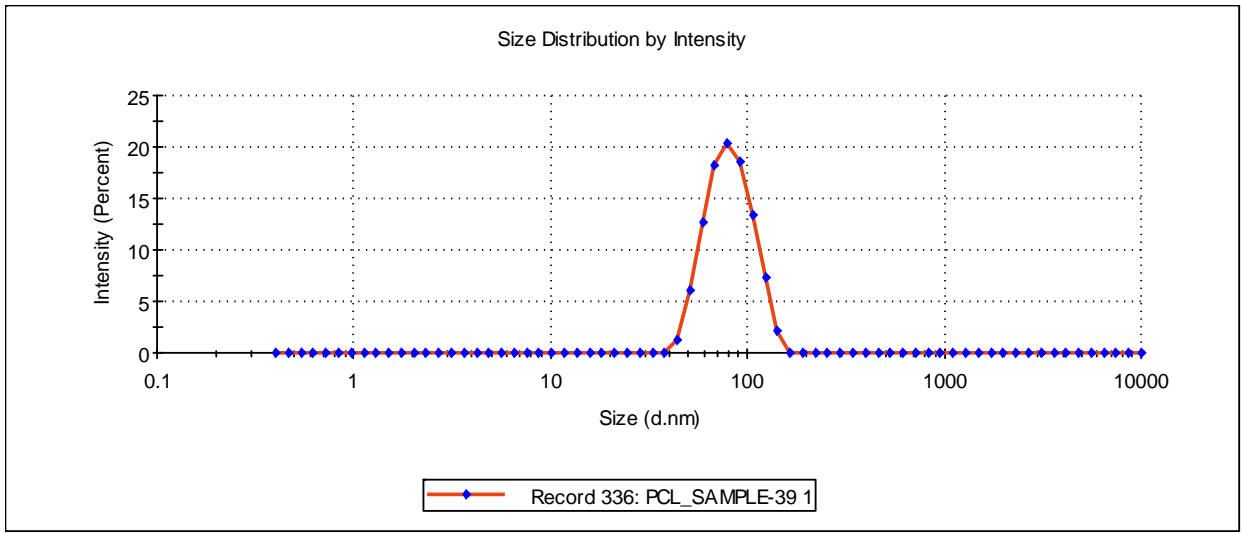

**Figure-3: Relationship between size distribution by percent intensity**

The PDI and PCL relationship in different experimental condition are shown in Figure 4. Except the 200 mg of PVA at 15 min of sonication time, the rest of the treatments 100 mg of PCL shows the PDI values close to zero, hence an indication of a homogenized solution. The treatment (PVA of 3500 mg at 15 min sonication time), PDI were close to zero for 100, 200 and 300 mg of PCL was used to encapsulate lycopene. However, the values were still high compared to the other treatments.

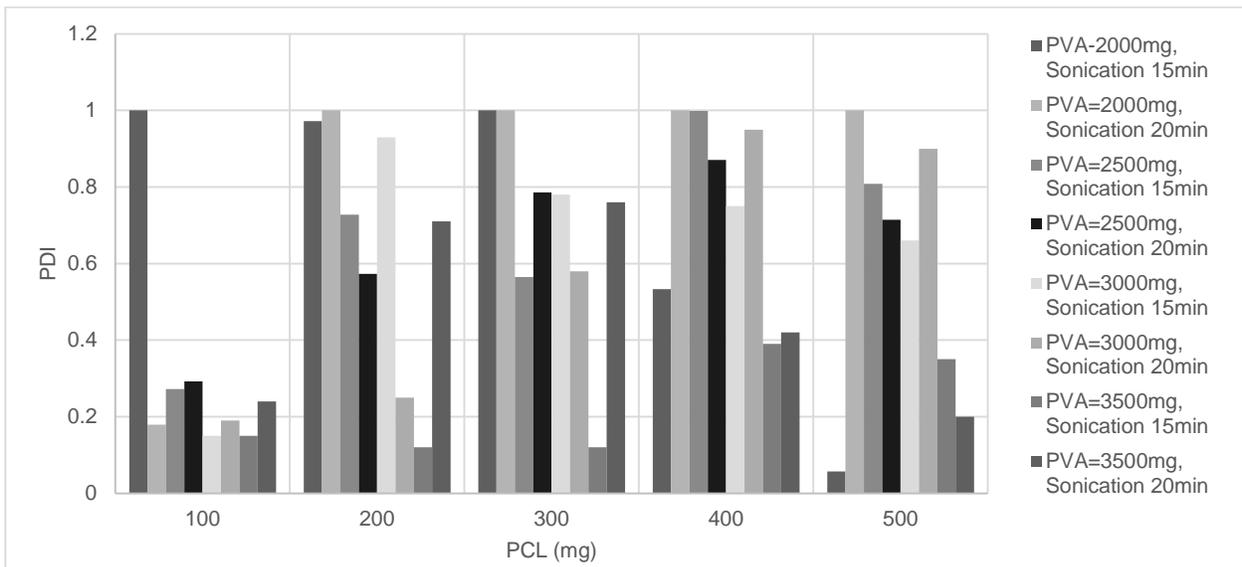

**Figure-4: The PDI profile of NP in different PCL concentrations**

Figure 5 shows that most of the value for all treatments have a PDI value below 0.20 or slightly above of it except when 500 mg of PLGA and 500 mg of DMAB were used coupled with 20 min of sonication.

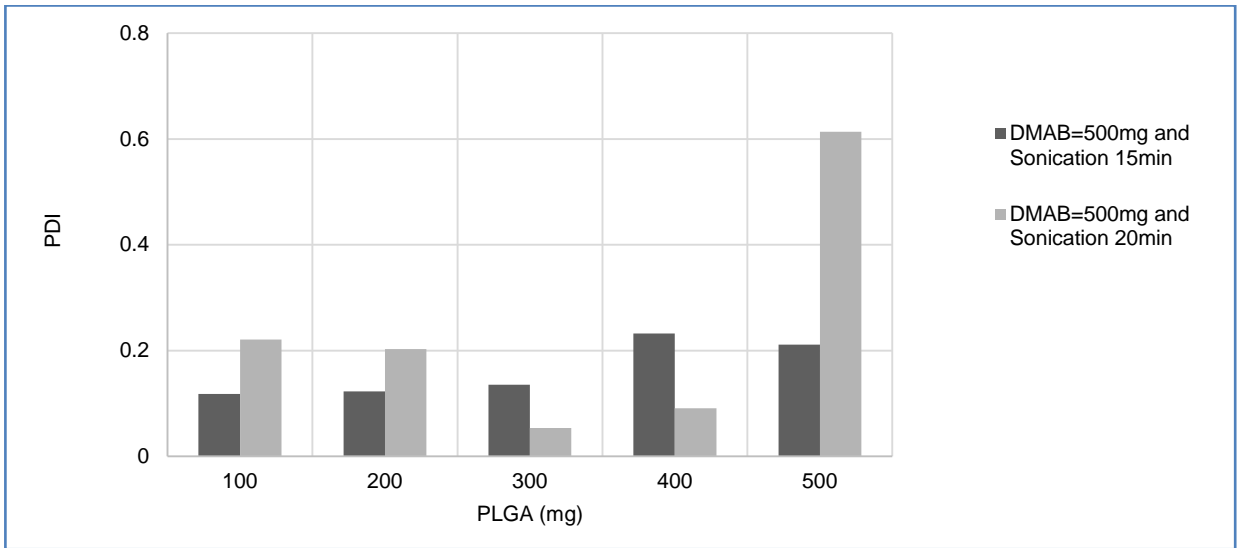

**Figure-5: The PDI profile of NP in different PLGA concentrations**

No linear relationship between the PCL concentration and zeta potential of nanoparticle solution at different treatments were found (Figure 6). In addition, the highest negative zeta potential value -4.21mV was found when 200 mg of PCL and 3500 mg of PVA was used to encapsulate lycopene.

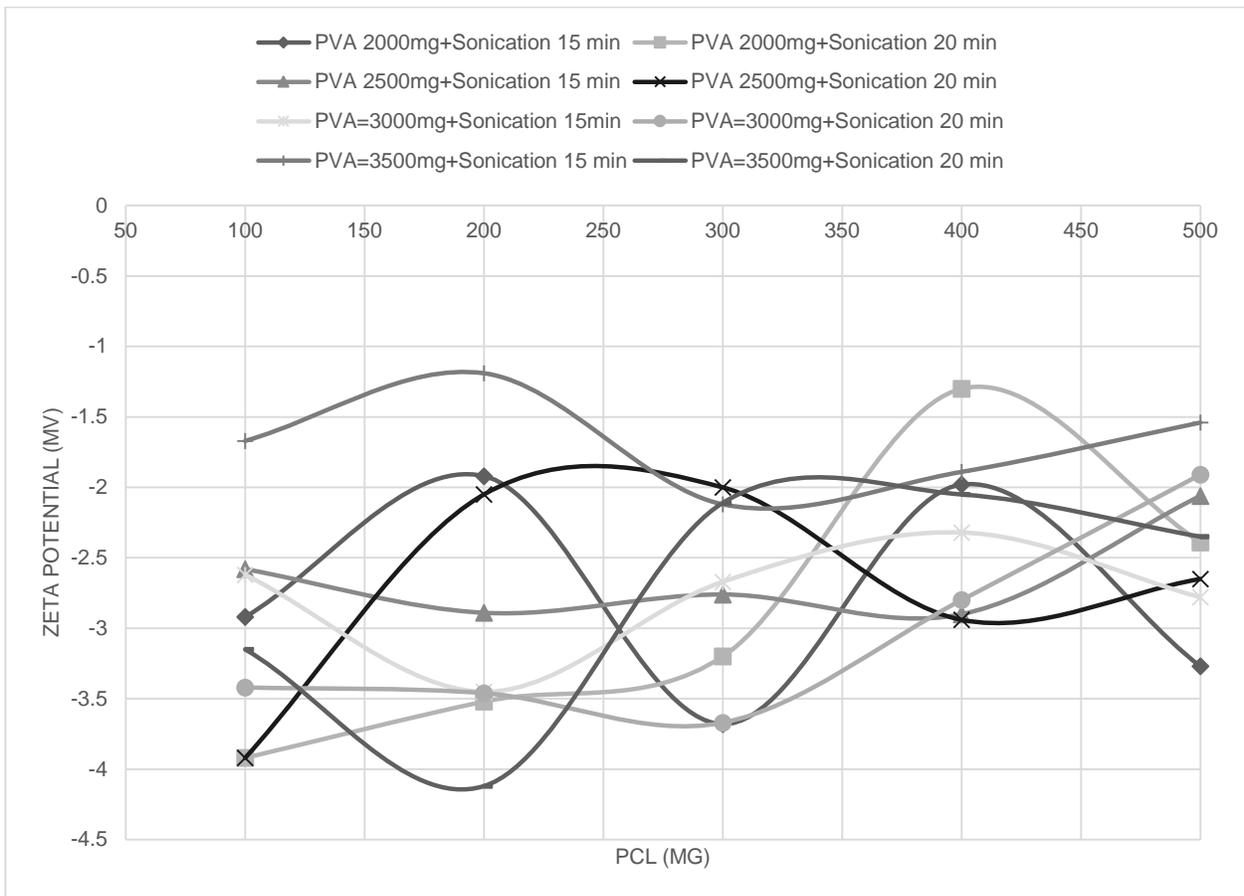

**Figure-6: The Zeta potential profile of NP in different PCL concentrations**

The higher the zeta potential higher the more stable is the NP solution. Figure 7 shows the values for zeta potential are very high, ranging from -72.36 ± 2.17 to 107.66 ± 3.15 mV which about 20 or more times greater than the NP solution synthesized with PCL.

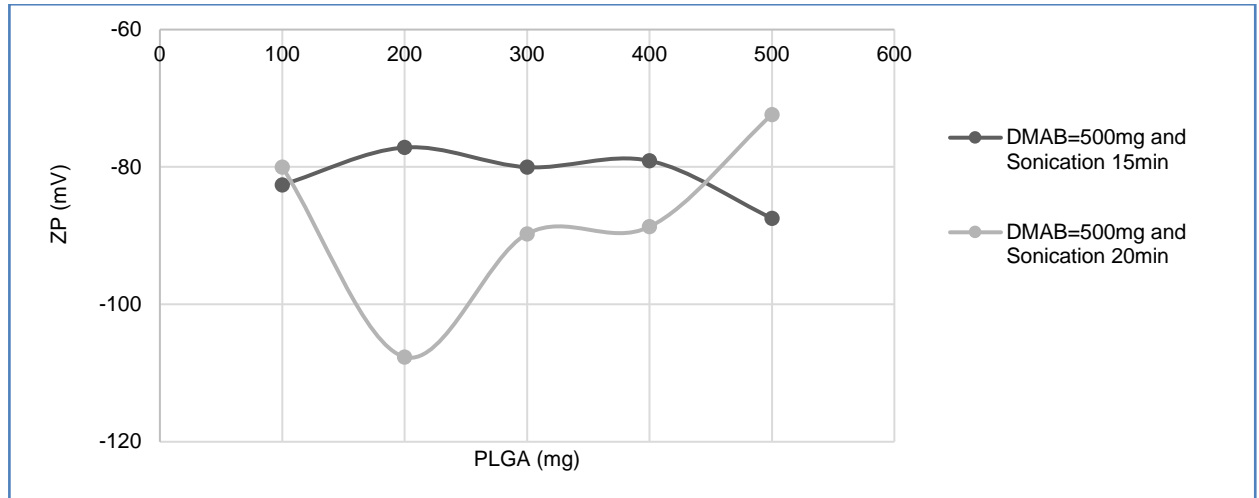

**Figure-7: The Zata potential profile of the NP in different PLGA concentrations**

## Conclusion:

The PCL nanoparticles prepared with emulsion evaporation technique showed smaller hydrodynamic diameter than PLGA. PDI value was similar for both the PCL and PLGA NP but the zeta potential found to be far too low with the PCL than the PLGA. There was no significant ($p < 0.05$) change on the hydrodynamic diameter, PDI and zeta potential values after 10 days storage of nanoparticle solution for both PCL and PLGA. As the erosion rate for PCL is lower than PLGA so PCL can be considered as a potential biodegradable polymer than PLGA to encapsulate lycopene for use in future control release studies.
.

## Acknowledgements

The authors wishes to acknowledge the USDA National Institute of Food and Agriculture, [USDA-NIFA Evans-Allen- Project Title: Develop Capacity in the Application of Nanotechnology in Active and Intelligent Food Packaging to Enhance Experiential Learning in Food Safety, Accession number: 1018224], and the Agriculture Experimental Research Station, Alabama A and M University for providing financial assistance to support this research project.
.